\documentclass[prl,showpacs,aps,twocolumn,floatfix]{revtex4}
\usepackage{amsmath}
\usepackage{graphicx}% Include figure files
\usepackage{dcolumn}% Align table columns on decimal point
\usepackage{bm}% bold math
\begin{document}

\title{Density matrix renormalization group approach for many-body open
quantum systems}

\author{J. Rotureau$^{1,2,3}$, N. Michel$^{1,2,3}$, W. Nazarewicz$^{1,2,4}$, M. P{\l}oszajczak$^5$, and J. Dukelsky$^6$}
\address{
$^1$ Department of Physics and Astronomy, University of Tennessee,
Knoxville, Tennessee
37996 \\
$^2$Physics Division, Oak Ridge National Laboratory, Oak Ridge, Tennessee 37831\\
$^3$Joint Institute for Heavy Ion Research,
Oak Ridge National Laboratory, P.O. Box 2008, Oak Ridge, Tennessee 37831\\
$^4$ Institute of Theoretical Physics, University of Warsaw, ul.
Ho\.za 69, 00-681 Warsaw, Poland\\
$^5$ Grand Acc\'el\'erateur National d'Ions Lourds (GANIL),
CEA/DSM - CNRS/IN2P3,
BP 55027, F-14076 Caen Cedex, France\\
$^6$ Instituto de Estructura de la Materia, CSIC, Serrano 123,
28006 Madrid, Spain}

\date{\today}

\begin{abstract}
The density matrix renormalization group (DMRG) approach is extended
 to complex-symmetric density matrices characteristic of  many-body open quantum
systems.  Within the continuum shell model, we investigate the
interplay between many-body configuration interaction  and
coupling to open channels. It is shown that the DMRG technique
applied to broad resonances in  the unbound neutron-rich
nucleus $^{7}$He provides  a highly accurate treatment of the
coupling to the non-resonant scattering continuum.
\end{abstract}


\maketitle

The DMRG method was introduced to overcome the limitations of 
Wilson-type renormalization group to describe strongly
correlated 1D lattice  systems with short range interactions
\cite{dmrg1} (see recent reviews  \cite{Rev1,Rev2}). Extensions of
DMRG to 2D lattice models are in general less accurate and require
a reduction to a 1D problem by selecting a path in the 2D plane.
More recently, by reformulating the DMRG in a single-particle
(s.p.) basis, several applications  to finite Fermi systems like
 molecules \cite{pap2}, superconducting grains \cite{pap3},
quantum dots \cite{pap4}, and atomic nuclei \cite{pap1} have been
reported. The main advantage of DMRG is that it can accurately
treat configuration spaces
 of dimensions  well beyond the limits of a large-scale diagonalization.
The early application of  DMRG  in the context of the nuclear
shell model (SM) in the $M$-scheme   exhibits
  some convergence problems that are not
yet understood \cite{pap5}. The first DMRG+SM applications in the
angular-momentum-conserving $J$-scheme were successfully carried
out in Ref.~\cite{revista} and also recently in
Ref.~\cite{pitsan}. A similar formalism, known as non-Abelian
DMRG, has been previously applied to quantum lattice models
\cite{nadmrg}.
 While most of the DMRG studies were focused on equilibrium
properties in strongly correlated closed quantum systems  with
hermitian density matrix,  non-equilibrium systems involving
non-hermitian and non-symmetric density matrices
 can also be treated
with a DMRG procedure \cite{pap6}. In this work, we shall present
the first detailed test  of the DMRG approach to many-body open
quantum systems (OQS)  described by the Gamow Shell Model (GSM)
\cite{Mic02,Bet02}, i.e., the continuum SM in the complex
$k$-plane.  The DMRG approach within the GSM is characterized
by complex-symmetric reduced density matrices. We are
particularly interested in developing an efficient strategy for
selecting most important configurations involving s.p. resonances
and non-resonant scattering states that describe many-body unbound
states. Our approach was briefly sketched  in Ref.~\cite{revista}.

GSM is the multi-configurational SM
with a s.p. basis given by the
Berggren ensemble \cite{Berggren} consisting of Gamow (resonant or Siegert)
states and the  non-resonant continuum
of scattering states. The resonant states are the generalized
eigenstates of the time-independent Schr\"{o}dinger equation,
 which are
regular at the origin and satisfy purely outgoing boundary conditions.
GSM provides a quasi-stationary formalism for describing the
time-dependent processes such as the multi-particle decays. The s.p.
Berggren basis is generated by a finite-depth potential, and the
many-body states can be expanded in
Slater determinants spanned by resonant and non-resonant
s.p. basis states \cite{Mic02,Bet02}.
The GSM, which has been up to now applied to weakly bound/unbound atomic
nuclei, can be equally applied for the description of other OQS, e.g.,
open microwave
resonators and  self-bound atomic systems, such as  $^3$He$_N$
neutral droplets at the limits of their stability  \cite{droplets}.
Another possible application  of the GSM are quantum dots, where
the interplay between electron-electron correlations and the continuum
coupling yields the transition in the conductance properties.

At present, the principal limitation of GSM applications is the
explosive growth in the number of configurations (i.e., dimension
of the many-body Fock space)  with both the number of active
particles and the size of the s.p. space. To ensure completeness
of the Berggren basis, for each resonant s.p. state of quantum
numbers $l_j$, with $l$  being the orbital angular momentum and $j$
the total angular momentum,  one should include a large set
of discrete non-resonant continuum states $\{(l_j)^{(i)};
i=1,\dots,M\}_{\rm c}$  lying on a continuous contour
$L_+^{l_{j}}$ in the complex $k$-plane \cite{Mic02,Bet02}. These
continuum states  become new active shells in the many-body
framework of GSM and, because of their presence, the dimension of
the GSM Hamiltonian matrix grows extremely fast. This explosion of
the Hilbert space is much more severe than in the standard SM,
which deals solely with a limited number of discrete s.p. states.
The use of the Berggren ensemble implies a {\em
complex-symmetric} Hamiltonian matrix for the {\em hermitian}
Hamilton operator. Due to the large number of non-resonant
continuum shells whose properties vary smoothly along the contour,
the GSM Hamiltonian matrix is significantly denser than that of a
conventional  SM. As a result, matrix operations are more
time-consuming in GSM than in SM, and
 the Lanczos procedure for GSM eigenvectors \cite{Mic02} becomes
very slow.

Most of the configurations involving many particles in the
non-resonant continuum contribute very little to low-energy GSM
eigenfunctions calling for a smart selection of the most important
configurations. For that purpose, we propose a new algorithm based
on the DMRG procedure for finding the GSM eigenvalues. The main
idea is to gradually consider different s.p. shells  of the
discretized non-resonant continuum in the configuration space and
retain  only   $N_{\rm opt}$ optimal  states dictated by  the
eigenvalues of the density matrix with largest modulus. The procedure will be
illustrated using the example of many-body resonances in the
neutron-unbound nucleus $^7$He  described in terms of  $N_v$=3
active (valence) neutrons outside  the closed core of $^{4}$He.

The s.p. basis is generated by a Woods-Saxon (WS) potential with
the radius $R_0$=2 fm, the depth of the central potential $V_0$=47
MeV, the diffuseness $d$=0.65 fm, and the spin-orbit strength
$V_{\rm so}$=7.5 MeV \cite{Mic02}. This s.p. potential reproduces
experimental energies and widths of the s.p. resonances  $3/2_1^-$
(g.s.) and $1/2_1^-$ (first excited state) in $^5$He
at $E$=0.745$-$$i$0.32\,MeV and
$E$=2.13$-$$i$2.94\,MeV, respectively.  The
neutron valence space consists of
 the $0p_{1/2}$ and  $0p_{3/2}$
resonant states and the corresponding non-resonant states
$\{p_{1/2}\}_{\rm c}$, $\{p_{3/2}\}_{\rm c}$. The $L_+^{p_{1/2}}$
contour in the complex-$k$ plane (in the following, $k$ is
expressed in units of fm$^{-1}$) is defined  by a triangle with
vertices at $({\cal R}e(k), {\cal I}m(k))$= (0,0), (0.33,-0.33),
(0.5,0.0), and a segment along the ${\cal R}e(k)$-axis  from
(0.5,0) to (1.0,0.0). Similarly, $L_+^{p_{3/2}}$ is given by a
triangle: (0,0), (0.17,-0.17), (0.5,0), and a straight segment
from (0.5,0) to (1.0,0). Each segment of these contours is
discretized with the same number of points corresponding to the
abscissas  for Gauss-Legendre quadrature. The GSM Hamiltonian is a
sum of the WS potential, representing the effect of an inert
$^4$He core, and the two-body interaction among valence neutrons.
The latter is approximated
 by a finite-range surface Gaussian interaction  \cite{Mic04}
  with  the range $\mu$=1\,fm
and the coupling constants depending on
the total angular momentum $J$ of the neutron pair:
$V_0^{(J=0)}$=$-542$ MeV fm$^{3}$, $V_0^{(J=2)}$=$-479$
MeV fm$^{3}$. These constants are fitted to
reproduce the  binding energies
of $^6$He and $^7$He with respect to the core.

For the purpose of DMRG, the configuration space is divided into
two Fock subspaces: $A$ (built from the s.p. resonant shells
$0p_{1/2}$, $0p_{3/2}$), and $B$ (built from s.p. non-resonant
shells $\{p_{1/2}\}_{\rm c}$ and $\{p_{3/2}\}_{\rm c}$). In the
warm-up phase,  one calculates and stores all possible matrix
elements of suboperators of the Hamiltonian in $A$: $a^{\dagger },
(a^{\dagger }\, \widetilde{a})^{K}, (a^{\dagger }a^{\dagger
})^{K}, ((a^{\dagger }a^{\dagger })^{K} \widetilde{a})^{L}),
((a^{\dagger }a^{\dagger })^{K}
(\widetilde{a}\widetilde{a})^{K})),$ and constructs all the states
$|k\rangle_A$  with $n=0,1,\cdots,N_v$ particles coupled to all
possible $j$-values. Then, from the sets: $\{ p_{1/2} \}_{\rm c}$,
$\{ p_{1/2} \}_{\rm c}$,  one picks up the first pair of s.p.
states, calculates the full set of matrix elements of suboperators
for this added pair, and constructs all the states $|i\rangle$
with $n$ particles ($n=0,1,\cdots,N_v$).  The subspace $B$ is
gradually enlarged by
 successively adding pairs
of non-resonant shells ($p_{1/2}, p_{3/2}$)   until
the number of states $|i\rangle_B$  exceeds  $N_{\rm opt}$. Then
the Hamiltonian is diagonalized in the space $\{
|k{\rangle}_A|i{\rangle}_B \}^{J}$ made of vectors in $A$ and $B$.
Obviously, the total number of particles in such states must be kept
 equal to the total
number of valence particles, and  the angular momentum
$J$ is conserved. To find the many-body resonance
from all the  eigenstates, $|\Psi^J
\rangle=\sum c_{ki}\{ |k{\rangle}_A|i{\rangle}_B \}^{J}$, one selects  the
eigenstate having the largest overlap with the  GSM eigenstate
in a pole approximation (i.e., diagonalized in the subspace $A$ only).
 From this eigenstate, one calculates the  reduced density matrix
$\rho ^{B,j}_{ii'}=\sum _{k}c_{ki}c_{ki'}$  with a fixed value
of $j$ in states $|i\rangle_{B}$, $|i'\rangle_{B}$ \cite {nadmrg}.
By construction, the density matrix $\rho ^{B}$ is then block-diagonal in $j$.
Within the metric defining the Berggren ensemble,
the density matrix is complex symmetric; hence,
its eigenvalues
are complex.
The  reduced  density matrix is
diagonalized, and $N_{\rm opt}$ eigenstates $|u_{\nu}\rangle$ having the
  eigenvalues with the  largest absolute values are retained.
 Those
eigenvalues correspond to the  most important states in the enlarged
set. All the matrix elements of suboperators for the optimized states
are recalculated.  Then, the next pair of non-resonant continuum shells is
added   and, again, only  $N_{\rm opt}$ eigenvalues
of the density matrix  are kept. This
procedure is repeated until the last shell in $B$ is reached, providing
a first guess for the wave function of the system  in  
the whole ensemble of shells forming the basis. From now on, all the s.p. states
are considered.
%//////////////////////////////////////////////////////////////////////////////
\begin{figure}[hbt]
\begin{center}
\includegraphics[width=6.5cm]{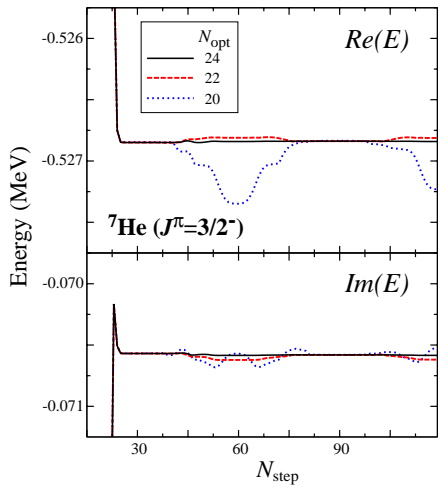}
\caption{GSM+DMRG results for the real  (top)
and imaginary (bottom) parts of the g.s. ($J^{\pi}=3/2^{-}$)
energy of $^{7}$He  plotted as a function of the number of DMRG steps.
The number of states kept during the sweeping phase
is $N_{\rm opt}$=20 (dotted line), 22 (dashed line), and
24 (solid line).
}
\label{fig1}
\end{center}
\end{figure}
%//////////////////////////////////////////////////////////////////////////////

At this point, the warm-up phase ends  and the so-called
sweeping phase begins.
Here,  one constructs states with $0,1,2,\cdots,N_v$ particles and then the
process  continues in the reverse direction until the number of vectors
becomes larger than
$N_{\rm opt}$.  If  the $m^{th}$  pair of $p$-shells in $B$ is reached, the Hamiltonian is diagonalized in the set of
vectors: $\{ |k, i_{prev}\rangle |i\rangle \}^{J}$,  where $i_{prev}$ is
a previously optimized state (constructed from
first $m-1$ pairs of  shells in $B$), and $i$ is a new  state. The density matrix is then diagonalized and  the $N_{\rm opt}$  $i$-states are kept. The procedure
continues by  adding the $(m-1)^{st}$ pair of shells, etc., until  the
first pair of $p$-shells in $B$ is reached. Then the procedure is reversed again:
the first pair of shells is added, then the second, the third, etc.

In the   examples presented in this work, the
number of s.p. shells included in  $A$ and $B$ are $N_{\rm r}$=2 and
$N_{\rm c}=60$ (30 shells in each continuum: $\{p_{1/2}\}_{\rm c}$,
$\{p_{3/2}\}_{\rm c}$), respectively. In the warm-up phase,  8
vectors  are kept.
Figure~\ref{fig1} shows the convergence properties of the ground state
(g.s.) $J^{\pi}=3/2^{-}$ resonance energy in $^7$He plotted as a
function of the number of DMRG steps $N_{\rm step}$.
(As discussed earlier, one step
 corresponds to adding a  pair of  $\{p_{1/2}\}_{\rm c}$, $\{p_{3/2}\}_{\rm c}$
shells.) Converged results, manifested by a local plateau at the complex
energy
$E$=$E_0$, are found after  $N_{\rm step}\leq 30$ steps. Then,
with increasing $N_{\rm step}$,  one enters the unstable region
$(40 \leq N_{\rm }\leq 70)$. A second plateau, corresponding to
practically the  same energy  $E_0$ as the first plateau
(and fairly independent of $N_{\rm opt}$),  appears
at $N_{\rm step}\sim 90$. The ``plateau-unstable" sequence
repeats periodically. It is worth noting that the deviation from
${\cal R}e(E_0)$ can be both positive and  negative. This is
because the Hamiltonian matrix is not hermitian but
complex-symmetric; hence, the usual
 reasoning  based on the  Ritz variational principle does not apply.

 The behavior of the DMRG procedure, as can be seen in
Fig.~\ref{fig1} and Table~\ref{table1}, shows a gradual reduction
of the energy variation around the plateau with increasing number
of sweeps $N_{\rm sw}$. Most importantly, it decreases quickly
 with the number of states $N_{\rm opt}$ kept in the sweeping
 phase. 
 The relative precision of  the DMRG procedure is
extremely high: $\sim$ $10^{-6}$ for
$N_{\rm opt}$=24 and $N_{\rm sw}$=4.
\begin{table}
\begin{center}
\begin{ruledtabular}
\begin{tabular}{|c|c|c|}
$N_{\rm opt}$ & $\Delta E/E$~~($N_{\rm sw}=2)$ & $\Delta E/E$~~($N_{\rm sw}=4)$   \\ \hline
$20$ & $9.4\times10^{-4}$ & $7.5\times10^{-4}$ \\
$22$ & $8.0\times10^{-5}$ & $4.9\times10^{-5}$ \\
$24$ & $3.0\times10^{-5}$ & $4.7\times10^{-6}$ \\
$26$ & $2.1\times10^{-5}$ & $2.6\times10^{-6}$
\end{tabular}
\end{ruledtabular}
\end{center}
\caption{\label{table1} Relative precision of the real part of the g.s. energy
in $^7$He as a function of the number of states $N_{\rm opt}$ kept in the
 sweeping phase at the second ($N_{\rm sw}$=2) and fourth ($N_{\rm sw}$=4)
 sweep of  GSM+DMRG.}
\end{table}

%//////////////////////////////////////////////////////////////////////////////
\begin{figure}[hbt]
\begin{center}
\includegraphics[width=6.5cm]{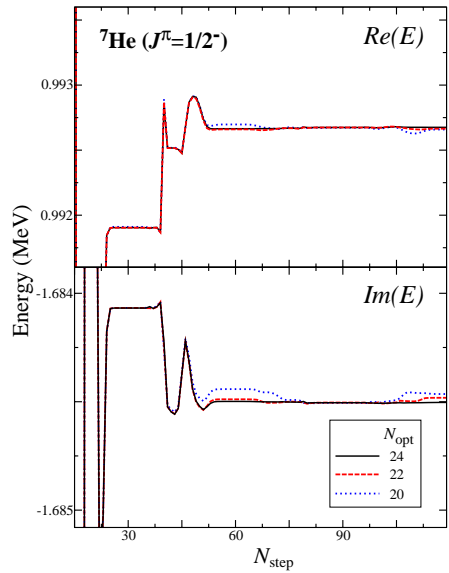}
\caption{Similar to Fig.~\ref{fig1} except for
 the first excited state ($J^{\pi}=1/2^-$)  in $^{7}$He.
}
\label{fig2}
\end{center}
\end{figure}
%//////////////////////////////////////////////////////////////////////////////
Figure~\ref{fig2} illustrates
the  convergence of the DMRG procedure for the first excited state in
$^7$He, which is calculated to be a broad resonance with
$\Gamma$=3.37 MeV. Here,
7 vectors  are kept in the warm-up phase.
 The full convergence is attained in about 2 sweeps, i.e., somewhat slower
than for the narrow g.s. resonance ($\Gamma$=0.14 MeV).
Still, the resulting plateau is excellent.  Again, as in
Fig.~\ref{fig1}, $E_0$ is practically independent of  $N_{\rm opt}$.
These two examples
demonstrate that the proposed DMRG algorithm  is
extremely efficient in selecting the most important non-resonant continuum
structures.

%//////////////////////////////////////////////////////////////////////////////
\begin{figure}[hbt]
\begin{center}
\includegraphics[width=6cm]{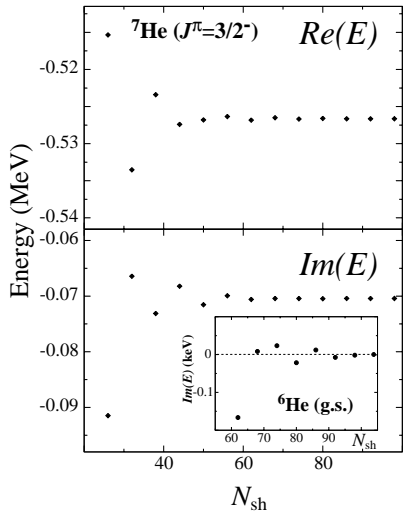}
\caption{Convergence of the real (top) and imaginary (bottom)
parts of the g.s. energy of $^7$He as a function of the the total
number of shells
$N_{\rm sh} (\equiv N_{\rm c}+N_{\rm r}=N_{\rm c}+2)$ included in the s.p. basis.
 As in Fig.~\ref{fig1}, 8 vectors
are kept in the warm-up phase. In the sweeping phase,
 $N_{\rm opt}$=22 states are kept. The inset shows the  imaginary part of the
 g.s. energy of $^6$He as a functiuon of $N_{\rm sh}$. 
 Here, $N_{\rm opt}$=4 (10)
 vectors were kept in the warm-up (sweeping) phase. 
}
\label{fig3}
\end{center}
\end{figure}
%//////////////////////////////////////////////////////////////////////////////
A major problem in  studies of OQS is to maintain the
completeness of the many-body
basis containing contributions from  {\em discretized}
non-resonant s.p.  continua. In this respect, the
Berggren ensemble is not different:
it  has to be discretized for  any
practical applications of the GSM. The accuracy of discretized many-body
calculations relies upon the independence of calculated observables
such as  energies,  transition probabilities, and cross sections
from the number  of s.p. shells  in the complex non-resonant continuum. 
 To illustrate this point, we show in Fig. \ref{fig3}
the g.s. energy of $^7$He  as a function of the total number of s.p. 
shells $\rm{N_{sh}}$.
One can see that the dependence of ${\cal R}e(E)$ and ${\cal
I}m(E)$ on $N_{\rm sh}$  is not monotonic, showing oscillations which
disappear only if a sufficiently dense discretization
is applied. A similar 
oscillatory pattern is seen
in the  imaginary part  of the g.s energy of $^6$He (shown in the inset).
 Since $^6$He is bound, the deviation of ${\cal I}m(E)$ from
zero is due to too coarse discretization; the completeness
is practically achieved for $N_{\rm sh}$=70. While the error due to an
insufficient number of non-resonant shells is of the order of $10^{-2}$\,keV, the 
accuracy of DMRG is in the range of $10^{-6}$\,keV.

The results shown in Figs.~\ref{fig1}-\ref{fig3} demonstrate that
the  fully
converged GSM results with respect to both the number of sweeps
and the number of shells in the
discretized continua can be obtained using the GSM+DMRG algorithm. The
rank of the biggest matrix to be diagonalized in GSM+DMRG grows
extremely slowly with  $N_{\rm sh}$. In the example shown in Fig.
\ref{fig3}, it  changes from $d$=941 to $d$=1001, whereas the dimension of
the GSM matrix in the  $J$-coupling scheme varies as $D\propto N_{\rm sh}^3$, from
$D$=6149 for $N_{\rm sh}$=26, to $D$=332,171 for $N_{\rm sh}$=98. Hence,
the gain factor $D(N_{\rm c})/d(N_{\rm c})\sim N_{\rm sh}^3$ quickly
grows  with increasing $N_{\rm c}$ (or the size of block $B$). 

Our results indicate that (i) 
the complex energy  $E_0$  at the DMRG plateau very weakly depends on $N_{\rm opt}$,
and (ii) the precision of calculated complex energies
 weakly depends  on  $N_{\rm sh}$, provided that  $N_{\rm sh}$ is sufficiently large. 
 This means that the
convergence features of the GSM+DMRG procedure can be tested by 
varying $N_{\rm sh}$ and $N_{\rm opt}$.
Once those parameters  are optimized in  `small-scale' GSM+DMRG
calculations, the  final  calculations can be performed in the large
model space to obtain fully converged results.

In summary,  this study describes the first application of the
DMRG method to unbound many-fermion systems  described by a
complex-symmetric eigenvalue problem. The encouraging features of
the proposed GSM+DMRG approach  open the scope to systematic
 and high-precision studies of the continuum coupling effects
in multiparticle OQS such as open microwave resonators,
multi-electron open quantum dots,  atomic nuclei, and atomic
clusters close to the particle drip line.

This work was supported by  the U.S. Department of Energy under
Contracts Nos. DE-FG02-96ER40963 (University of Tennessee),
DE-AC05-00OR22725 with UT-Battelle, LLC (Oak Ridge National
Laboratory), and DE-FG05-87ER40361 (Joint Institute for Heavy Ion
Research), and by the Spanish DGI under grant No.
BFM2003-05316-c02-02.

%%%%%%%%%%%

\end{document}